\documentclass[a4paper]{jpconf}
\usepackage{graphicx,epsfig}
\usepackage{amsmath,amssymb}
\usepackage{arydshln}

\newcommand{\ord}[1]{\mathcal{O}\left({#1}\right)}
\newcommand{\cO}{{\cal O}}

\begin{document}

%opening
\title{RGE behaviour of $U(2)^3$ symmetry in SUSY}
\author{J Jones P\'erez}
\address{
Departament de F\'{\i}sica Te\`orica and IFIC, Universitat de Val\`encia-CSIC, \\
E-46100, Burjassot, Spain}
\ead{joel.jones@uv.es}

\begin{abstract}
From the naturalness point of view, the first LHC results seem to disfavour any constrained MSSM realization with universal conditions at the SUSY-breaking scale. A more motivated scenario is given by split-family SUSY, in which the first two generations of squarks are heavy, compatible with a $U(2)^3$ flavour symmetry. Here, after reviewing the flavour structures obtained in this framework, we consider the flavour symmetry to be broken at a very high scale, and study the consequences at low energies through its RGE evolution. Initial conditions compatible with a split scenario are found, and the preservation of correlations from minimal $U(2)^3$ breaking are checked.
\end{abstract}

\section{Introduction}

When considering physics beyond the Standard Model (SM), Supersymmetry, in the form of the Minimal Supersymmetric Standard Model (MSSM) is usually considered as one of the most appealing options. However, due to its vast parameter space, a standard procedure is to assume universality conditions of the soft terms at the GUT scale, and allow the renormalization group equations (RGEs) to break the universality when running down to the weak scale. This constrained realization is named CMSSM, and leads to flavoured soft terms acquiring a structure following Minimal Flavour Violation (MFV)~\cite{D'Ambrosio:2002ex}.

MFV effectively suppresses all SUSY contributions to flavoured processes, solving the so-called SUSY flavour problem. However, it also leads to sfermion masses with small splittings. This can cause issues in the solution of the hierarchy problem, which depends mainly on having light third generation squark masses. The issue is caused due to the lack of SUSY signals at the LHC, which forces the masses of the first two generation squarks to be pushed to larger values. The MFV structure causes the first two generations to drag with them the mass of the third generation squarks to these larger values, compromising in this way the solution of the hierarchy problem.\footnote{One must admit that the observed value of the Higgs mass already spoils the solution of the hierarchy problem in most of the SUSY parameter space. Nevertheless, a lower value of the stop mass can effectively reduce the fine-tuning, even though it might not be able to avoid it entirely.}

One would therefore like to have a framework similar to MFV, such that contributions of flavoured processes are still suppressed to some degree, but allowing the third generation masses to be decoupled from the first two. One would also hope that the flavour suppression in this new framework be somewhat milder, such that some significant effects could be observed in flavour experiments. In particular, one would hope these new effects could solve the ongoing tension between CP violating $\Delta F=2$ observables in the $K$ and $B_d$ sectors~\cite{Altmannshofer:2009ne}.

A framework based on a $U(2)^3$ symmetry in the quark sector was built in~\cite{Barbieri:2011ci}, and was found to satisfy all of the above requirements. Further studies of this symmetry were carried out in~\cite{Barbieri:2011fc}, where the squark and slepton phenomenology was analysed in further detail, both in and out of SUSY. Nevertheless, in these works the symmetry structure was applied directly at the electroweak scale, with no consideration of RGE running. As one would expect that the breaking of such a symmetry would occur at a high scale, it was found necessary to study the consequences of the running on the low-scale structures. In particular, it was unclear if the virtues of $U(2)^3$ would be maintained at the weak scale, and what kind of initial conditions could lead to the required split scenario.

The current paper is a summary of the work carried out in~\cite{Blankenburg:2012ah}. In Section~\ref{sec:rev} we briefly review the $U(2)^3$ framework and the conditions leading to the solution of the flavour tension. In Section~\ref{sec:rge} we show the result of our study of the RGE evolution, and then conclude.

\section{Review of $U(2)^3$ Framework in SUSY}
\label{sec:rev}

We shall restrict ourselves to the quark sector. In contrast to MFV, which assigns a $U(3)$ symmetry for each flavoured chiral supermultiplet, this framework assigns a $U(2)$ symmetry for each supermultiplet, under which only the first two generations shall transform. In the limit of exact flavour symmetry, the only terms allowed in the superpotential involving flavoured fields are those of the third generation (s)quarks:
\begin{equation}
 W= y_t\,Q_3 t_R^c H_u - y_b\,Q_3 b_R^c H_d~.
\end{equation}
As in MFV, we proceed to break the flavour symmetries by adding spurion fields. The minimal way of constructing the Yukawas is by adding bi-doublet spurions $\Delta Y_u$ and $\Delta Y_d$, from which we build the structure of the $(1-2)$ block, and a doublet $V$, transforming only under $U(2)_Q$, used to connect the first two generations with the third. With these assumptions, the Yukawas acquire the structure:
\begin{align}
Y_u= y_t \left(\begin{array}{c:c}
 \Delta Y_u & x_t\,V \\ \hdashline
 0 & 1
\end{array}\right), & &
Y_d= y_b \left(\begin{array}{c:c}
 \Delta Y_d & x_b\,V \\ \hdashline
 0 & 1
\end{array}\right)~,
\end{align}
with everything above the horizontal dashed line having two rows, and everything to the left of the vertical dashed line having two columns. We find it possible to parametrize the whole matrices in terms of the eigenvalues of the bi-doublets, $\lambda_{f_1}$ and $\lambda_{f_2}$, a left-mixing matrix with one angle and phase per bi-doublet, $s_f$ and $\alpha_f$, a suppression parameter $\epsilon$ for the doublet, and two complex couplings $x_f e^{i\phi_f}$. Notice only three phases are independent.

With this setup, one can build the CKM matrix, and fit the following parameters~\cite{Barbieri:2011ci}:
\begin{align}
 s_u&=0.095\pm0.008~, & s_d&=-0.22\pm0.01~, \nonumber \\
 s&=0.0411\pm0.0005~, & \cos(\alpha_u-\alpha_d)&=-0.13\pm0.2~,
\end{align}
where $s\propto\epsilon$. We choose to fix $\epsilon=\lambda_{\rm CKM}^2$.
 
After determining which spurions are needed to build the Yukawa matrices, we follow the MFV assumption demanding all flavour structures to depend exclusively on these spurions. With this, we find that the soft terms acquire the following form:
\begin{eqnarray}
\label{mq.spur}
\frac{m^2_{\tilde Q}}{m^2_{Q_h}} &=&  I+
\left(\begin{array}{c:c} 
c_{Qv}\,V^* V^T   +  c_{Qu}  \Delta Y_u^* \Delta Y_u^{T} 
+  c_{Qd}  \Delta Y_d^* \Delta Y_d^{T} 
&   x_{Q}\, e^{-i\phi_Q} V^* \\ \hdashline
    x_{Q}\, e^{i\phi_Q} V^T   &   -\rho_Q 
\end{array}\right)~,   \\
\frac{m^2_{\tilde d}}{m_{d_h}^2} &=&  I+
\left(\begin{array}{c:c}
c_{dd}\,\Delta Y_d^\dagger \Delta Y_d
&  x_{d}\, e^{-i\phi_d} \Delta Y_d^\dagger V \\ \hdashline
   x_{d}\, e^{i\phi_d} V^\dagger \Delta Y_d  & -\rho_u
\end{array}\right)~,  \\
\frac{m^2_{\tilde u}}{m_{u_h}^2} &=& I+
\left(\begin{array}{c:c} 
c_{uu}\,\Delta Y_u^\dagger \Delta Y_u
&    x_{u}\, e^{-i\phi_u} \Delta Y_u^\dagger V \\ \hdashline
     x_{u}\, e^{i\phi_u} V^\dagger \Delta Y_u   & -\rho_d 
\end{array}\right)~,
\end{eqnarray}
where $\rho_f=(m^2_{f_h}-m^2_{f_l})/m^2_{f_h}$, and all $c_i$ and $x_i$ parameters are real, of $\ord{1}$.

These structures lead to the following contributions to $\Delta F=2$ observables:
\begin{align}
\epsilon_K&=\epsilon_K^\text{SM(tt)}\times\left(1+x^2F_0\right) +\epsilon_K^\text{SM(tc+cc)} ~
\label{eq:epsKxF}\\
S_{\psi K_S} &=\sin\left(2\beta + \text{arg}\left(1+xF_0 e^{-2i\gamma}\right)\right) ~,\label{eq:Spk} \\
\Delta M_d &=\Delta M_d^\text{SM}\times\left|1+xF_0 e^{-2i\gamma}\right| ~,
\label{eq:DMdxF}\\
\frac{\Delta M_d}{\Delta M_s} &= \frac{\Delta M_d^\text{SM}}{\Delta M_s^\text{SM}} ~,
\label{eq:MdMs}
\end{align}
where:
\begin{eqnarray}
F_0 &=& \frac{2}{3} \left(\frac{g_s}{g} \right)^4 \frac{ m_W^2 }{ m^2_{Q_3} } \frac{1}{S_0(x_t)}
\left[ f_{0}(x_g) + \cO\left( \frac{m_{Q_l}^2}{m_{Q_h}^2} \right)\right]~, 
\qquad x_g = \frac{ m^2_{\tilde g} }{ m_{Q_3}^2 }~,
\label{eq:F0}
\\
f_{0}(x_g) & = & \frac{11 + 8 x_g -19x_g^2 +26 x_g\log(x_g)+4x_g^2\log(x_g)}{3(1-x_g)^3}~, 
\qquad x=\frac{s_L^2 c^2_d}{|V_{ts}|^2}
\end{eqnarray}
with $S_0(x_t)$ being the typical one-loop function of the SM to $\Delta F=2$ processes. The $s_L\,e^{i\gamma}$ parameters are defined in~\cite{Barbieri:2011ci}, and are related to the $\epsilon$ and $\phi_f$ terms.

\begin{figure}
\begin{center}
\includegraphics[width=0.4\textwidth]{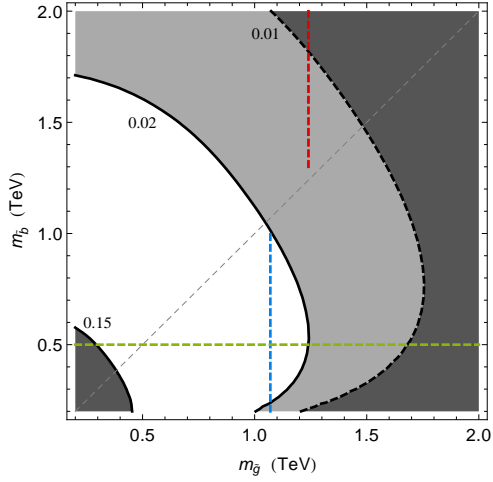}
\end{center}
\caption{Value of $F_0$ as a function of $m_{\tilde g}$ and $m_{\tilde b}$. White region has $0.02<F_0<0.15$, grey region has $F_0>0.01$. Red and blue lines are rough limits on $m_{\tilde g}$, taken from ATLAS, green line is a rough limit on $m_{\tilde b}$, taken from CMS.}
\label{fig:F0}
\end{figure}

With these modifications, one can find that the $U(2)^3$ framework can provide a large enough contribution to $\epsilon_K$ and $S_{\psi K_s}$ in the correct directions, such that the flavour tension mentioned above is solved. For not too large $x$, we required $F_0\gtrsim0.02$, but found that values of $F_0$ as low as $0.01$ could also reduce the tension by compensating with slightly large $\ord{1}$ parameters. This implies upper bounds on gluino and sbottom masses, as shown in Figure~\ref{fig:F0}.

Figure~\ref{fig:F0} also shows the current status of the $U(2)^3$ framework in SUSY. The red and blue lines indicate lower bounds on $m_{\tilde g}$ set by the ATLAS collaboration, for the cases $m_{\tilde g}<m_{\tilde b}$~\cite{ATLAS:2012qna} and $m_{\tilde g}>m_{\tilde b}$~\cite{Aad:2012pq}, respectively. The green line indicates lower bounds on $m_{\tilde b}$ by the CMS collaboration~\cite{Chatrchyan:2012wa}. The white area shows $0.02<F_0<0.15$, and is almost completely excluded by direct searches. The grey area has $0.01<F_0<0.02$, and is not heavily constrained when $m_{\tilde g}>m_{\tilde b}$. Thus, current data favours heavy gluinos, and somewhat large mixing parameters.

\section{RGE Behaviour of $U(2)^3$ Framework}
\label{sec:rge}

We now turn to study the properties of this framework when defined at a high energy scale, and then evolved down to the electroweak scale. We address three questions: {\bf (1)} What sort of initial conditions should we look for? {\bf (2)} How does the mixing evolve? {\bf (3)} Are the virtues of $U(2)^3$ preserved at all scales?

To simplify the study of the initial conditions, we shall turn to a CMSSM-like setup. At the GUT scale, we keep the universal gaugino mass, $M_{1/2}$, but differentiate between the common Higgs mass, $m_0$, the common (heavy) first two generations scalar mass, $m_{heavy}$, and the common (light) third generation scalar mass, $m_{light}$. Moreover, we use the parameter $\rho=(m_{heavy}^2-m_{light}^2)/m_{heavy}^2$ instead of $m_{light}$. The A-Terms follow a structure similar to that of the Yukawas, with different $\ord{1}$s and multiplied by a factor $A_0\sim m_{heavy}$. After running the parameters down to the electroweak scale, we require no tachyonic states, radiative electroweak symmetry breaking, and the correct Higgs mass within $1\sigma$. We also prefer regions where $F_0>0.01$, and attempt to reduce the hierarchy problem by having small values of $\mu$ and $m_{\tilde t}$.

\begin{figure}[tbp]
\begin{center}
\includegraphics[width=0.32\textwidth]{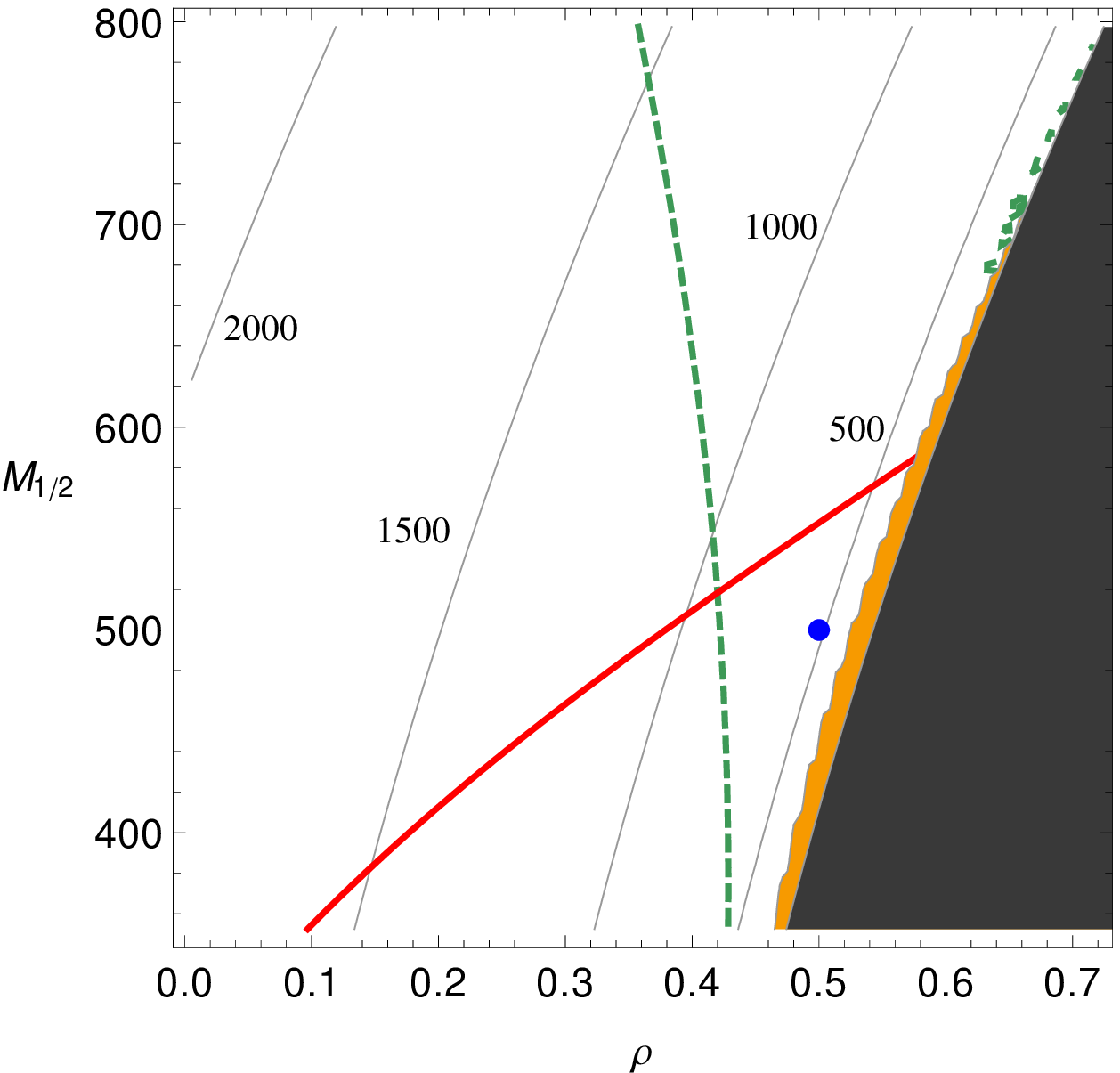}
\includegraphics[width=0.32\textwidth]{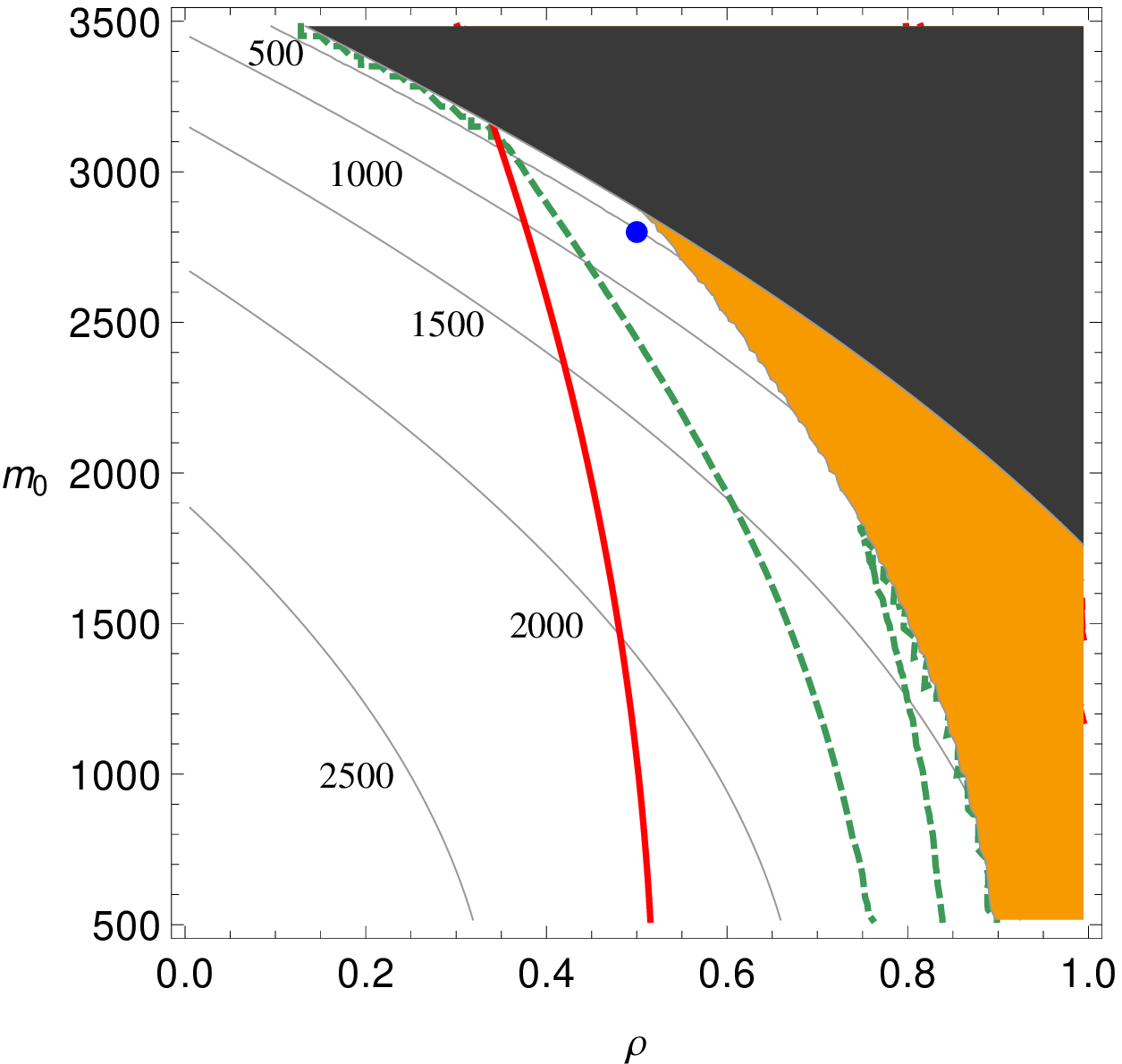}
\includegraphics[width=0.32\textwidth]{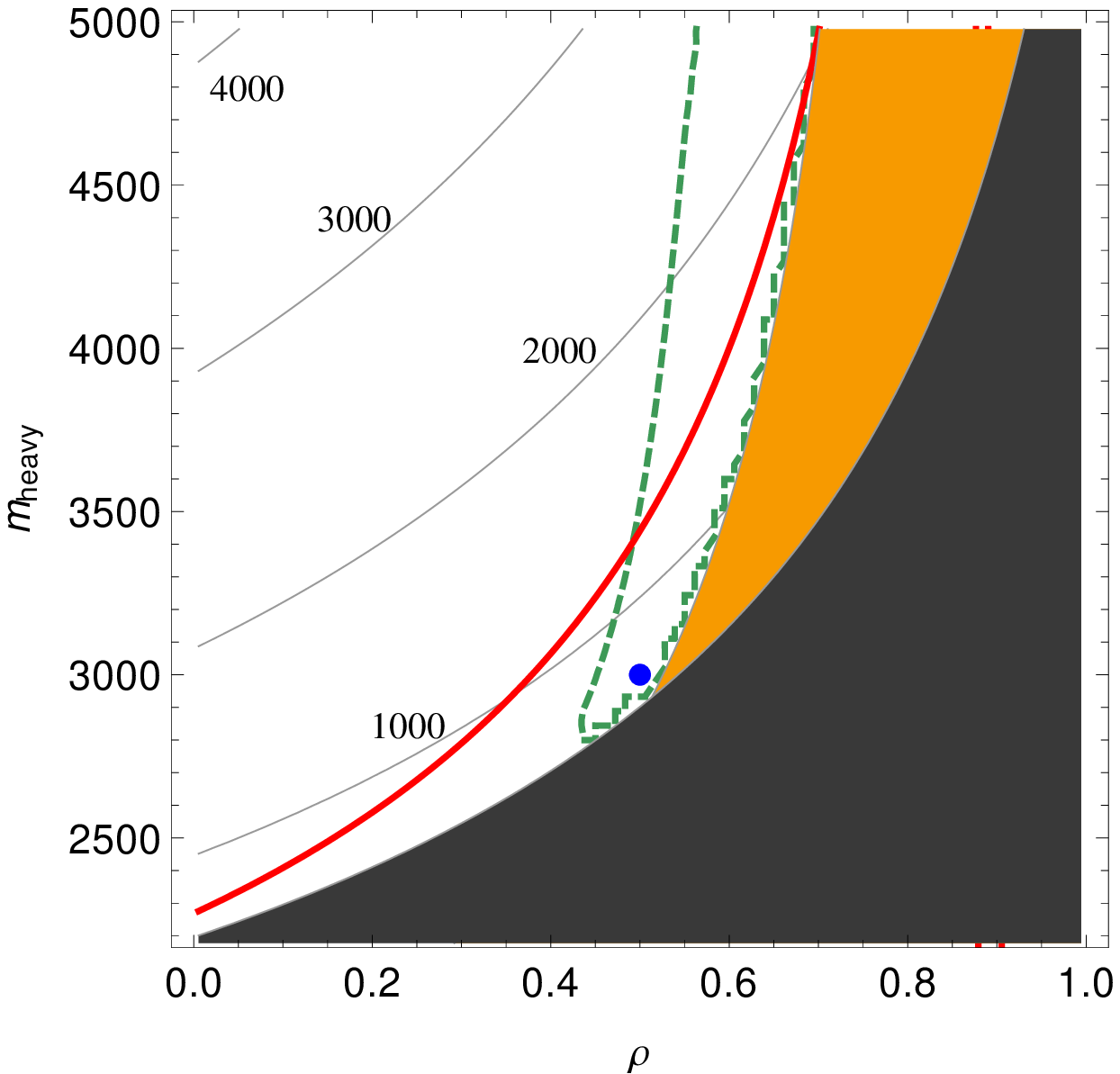} \\
\end{center}
\caption{Parameter space around our benchmark, with contours for $\mu$. The dark regions correspond to no EWSB, while the orange regions have a tachyonic stop. The green, dashed lines delimitate the regions within 1 sigma of the Higgs mass, and the red, solid curve indicates the regions below which the flavour tension could be solved. The blue dot represents the benchmark.} 
\label{fig:benchmark1}
\end{figure}

Although in~\cite{Blankenburg:2012ah} we studied two different benchmarks, here we shall concentrate only in one, and describe the parameter space around this point. The benchmark is characterized by $M_{1/2}=0.5$~TeV, $m_0=2.8$~TeV, $m_{heavy}=3$~TeV, $\rho=0.5$ and $A_0=-m_{heavy}$. We show in Figure~\ref{fig:benchmark1} the parameter space around this point, with the different constraints represented in different colours. We can see that the region that fulfills all constraints is very small, especially if we demand $\mu<1$~TeV.

The benchmark leads to values of $\mu$ around 500 GeV, and very specific splittings. In the stop sector, we find $\rho^{\rm low}_{\tilde t}\sim0.85$, which leads to an average stop mass of about $1.2$~TeV. Nevertheless, as the stop mixing is large, we find the mass of the lightest stop to be lower than 500~GeV. On the other hand, in the sbottom sector, we have $\rho^{\rm low}_{\tilde b}\sim0.6$, leading to an average sbottom mass of $1.9$~TeV. Notice that this setup involves a very mild splitting at the GUT scale, but can lead to a larger splitting in the stop sector.

Having found a point with suitable initial conditions, we now turn to the study of the mixing. To this end, we use objects directly related to physical observables, which are also connected with the mixing, and study their RGE evolution. As our interest lies on $\Delta F=2$ processes, which receive their main contribution from $(LL)^2$ operators, the objects in question are defined as \begin{equation}
\lambda^{(a)}_{i\not = j} = (W^d_L)_{ia} (W^d_L)^*_{ja}~,
\end{equation}
where $W^d_L$ is the diagonalization matrix of $m_{\tilde{Q}}^2$ in the basis of diagonal down quarks. In~\cite{Barbieri:2011ci}, it was shown that the $U(2)^3$ structure leads to $\lambda^{(3)}_{12}$, $\lambda^{(3)}_{13}$ and $\lambda^{(3)}_{23}$ to be closely related:
\begin{align}
\label{eq:lambda}
\lambda^{(3)}_{12}= s_L^2 \kappa^* c_d~, & & \lambda^{(3)}_{13}= -s_L \kappa^* e^{i\gamma}~, 
& & \lambda^{(3)}_{23}=-c_d s_L e^{i\gamma}~,
\end{align}
where $\kappa\approx c_d V_{td}/V_{ts}$. Thus, assuming for now that these relations are held throughout the running, we shall only need to study the evolution of one of these objects, which we choose to be $\lambda^{(3)}_{23}$.

\begin{figure}[tbp]
\begin{center}
\includegraphics[width=0.45\textwidth]{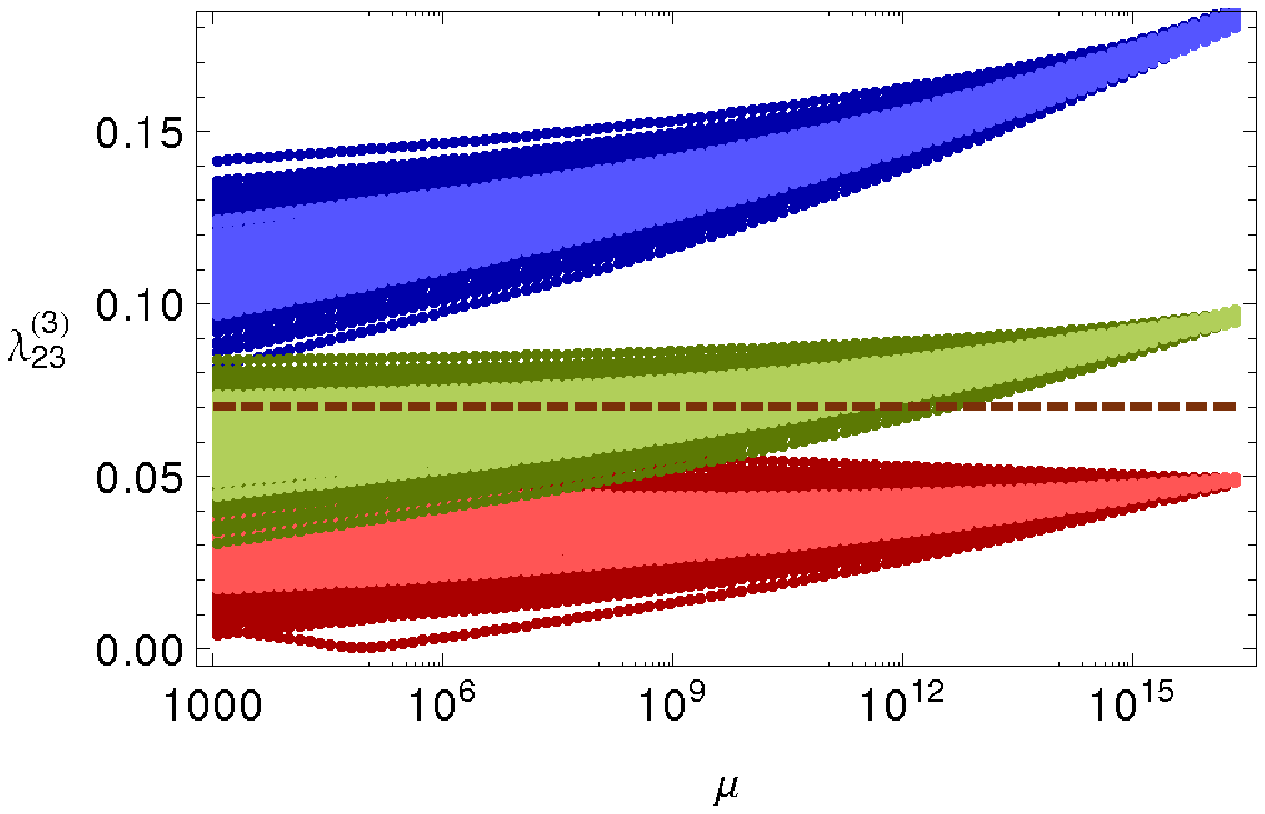} \qquad
\includegraphics[width=0.45\textwidth]{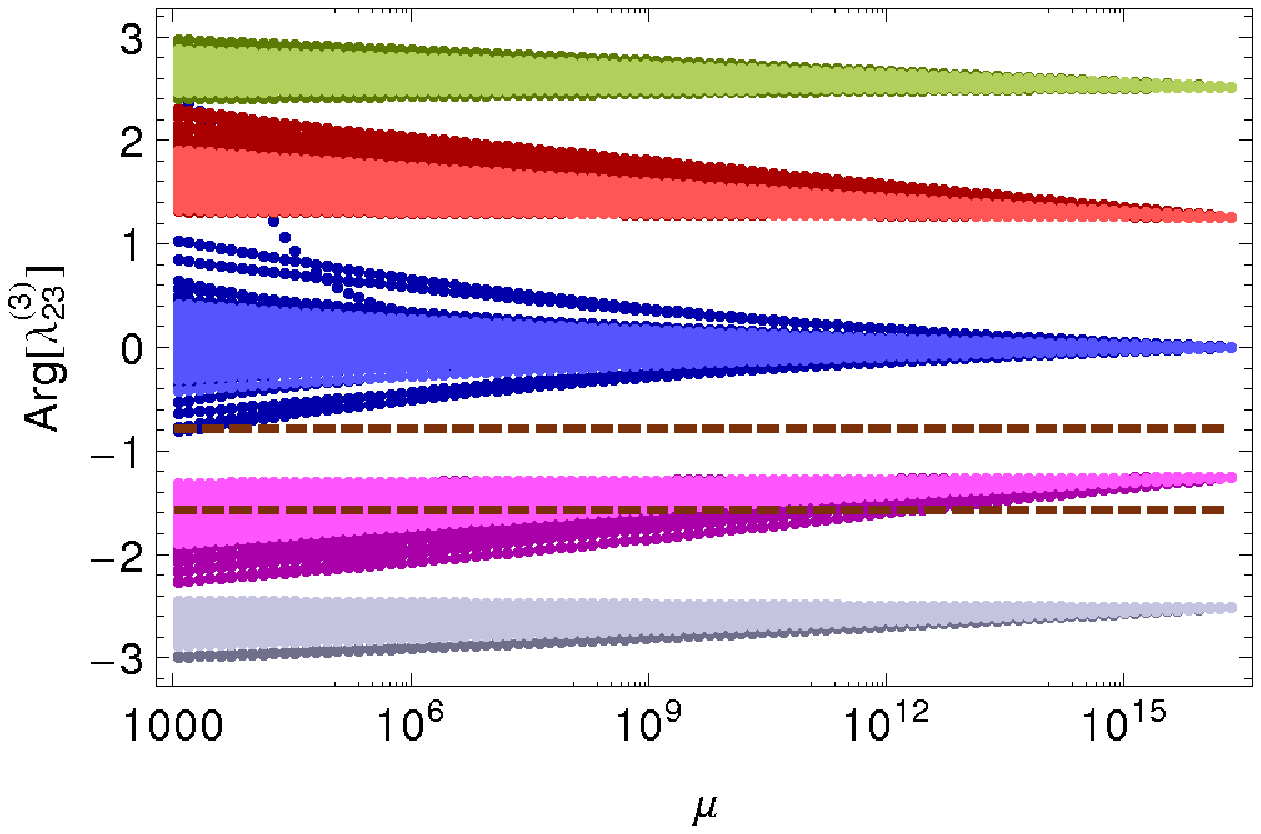} %\\
\end{center}
\caption{The running of $|\lambda_{23}^{(3)}|$ (left) and Arg$(\lambda_{23}^{(3)})$ (right). On the left, we show $x_L=2,\,1,\,0.5$ in blue, green and red. In every region the lighter colour correspond to $\gamma$ fixed to $\pi/4$. On the right we fix $\gamma=(-1+0.4n)\pi$, with $n=0,\,1,\,2,\,3,\,4$ in blue, red, green, grey and magenta, respectively. The lighter regions correspond to $x_L=1$. In the first two plots the dashed brown lines mark the region where the flavour tension can be solved. On the left, the region is above the line, while and on the right it is between the two lines.}
\label{fig:lambdasB1}
\end{figure}

In Figure~\ref{fig:lambdasB1} we show how $\lambda^{(3)}_{23}$ evolves with the scale. The main result is that, although there exists some suppression effect in the magnitude, the mixing remains relatively stable throughout the running. The Figure also shows the minimum value that the mixing must have in order to solve the flavour tension, according to~\cite{Barbieri:2011ci}. Here, we can see that the $\ord{1}$ parameters cannot be too small, which is a direct consequence of the small value of $F_0$.

Finally, we need to check if the $U(2)^3$ relations of Equation~(\ref{eq:lambda}) are kept after the RGE running. The best way to do this is to use the $\lambda^{a}_{ij}$ to define new objects that should be approximately RGE invariant, at least within their theoretical errors, and check if this invariance holds. We choose:
\begin{align}
 \frac{\lambda^{(3)}_{13}}{\lambda^{(3)}_{23}} = \frac{\kappa^*}{c_d}~, & & 
 \frac{\lambda^{(3)}_{12}}{|\lambda^{(3)}_{13}|^2} = \frac{c_d}{\kappa}~,
\end{align}
The first ratio tests the correlations between the $B_d$ and $B_s$ sectors, the second ratio tests those between the $K$ and $B_d$ sectors. If we find these ratios to hold within their theoretical errors, we can consider the $U(2)^3$ symmetry to be preserved by the running.

However, we first need to write the theoretical error for each of these quantities. The main source of error in the absolute value of each ratio lies on a NLO effect due to $\rho\neq1$, which can generate deviations of up to $4\%$ the value of the ratio. On the other hand, the phase has a fixed correction, of order $\varphi_c=\arg(c_u c_d+s_u s_d e^{-i\phi})\approx0.02$. We shall then take the error on the phase equal to $\varphi_c$, after including the correction. Moreover, we neglect the errors due to the running of the CKM elements in $\kappa$.

\begin{figure}[tb]
\begin{center}
\includegraphics[width=0.45\textwidth]{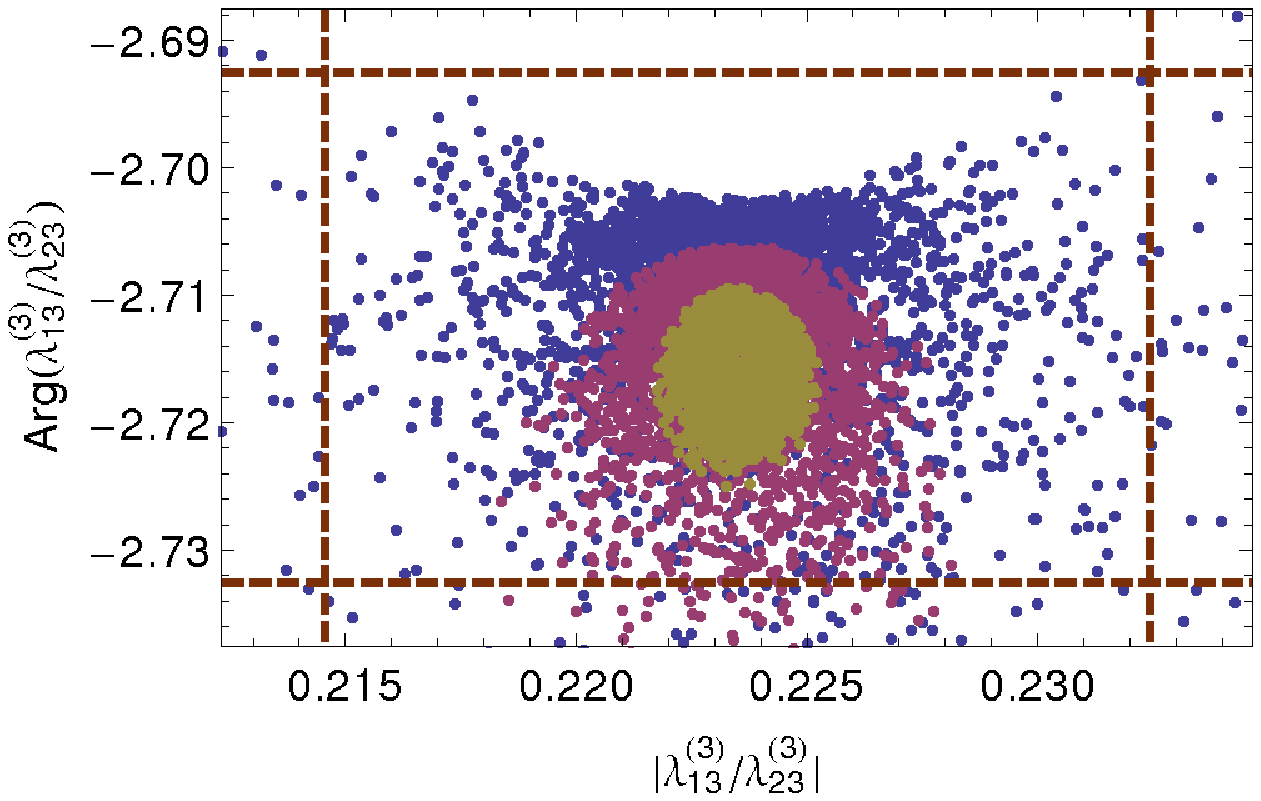} \qquad
\includegraphics[width=0.45\textwidth]{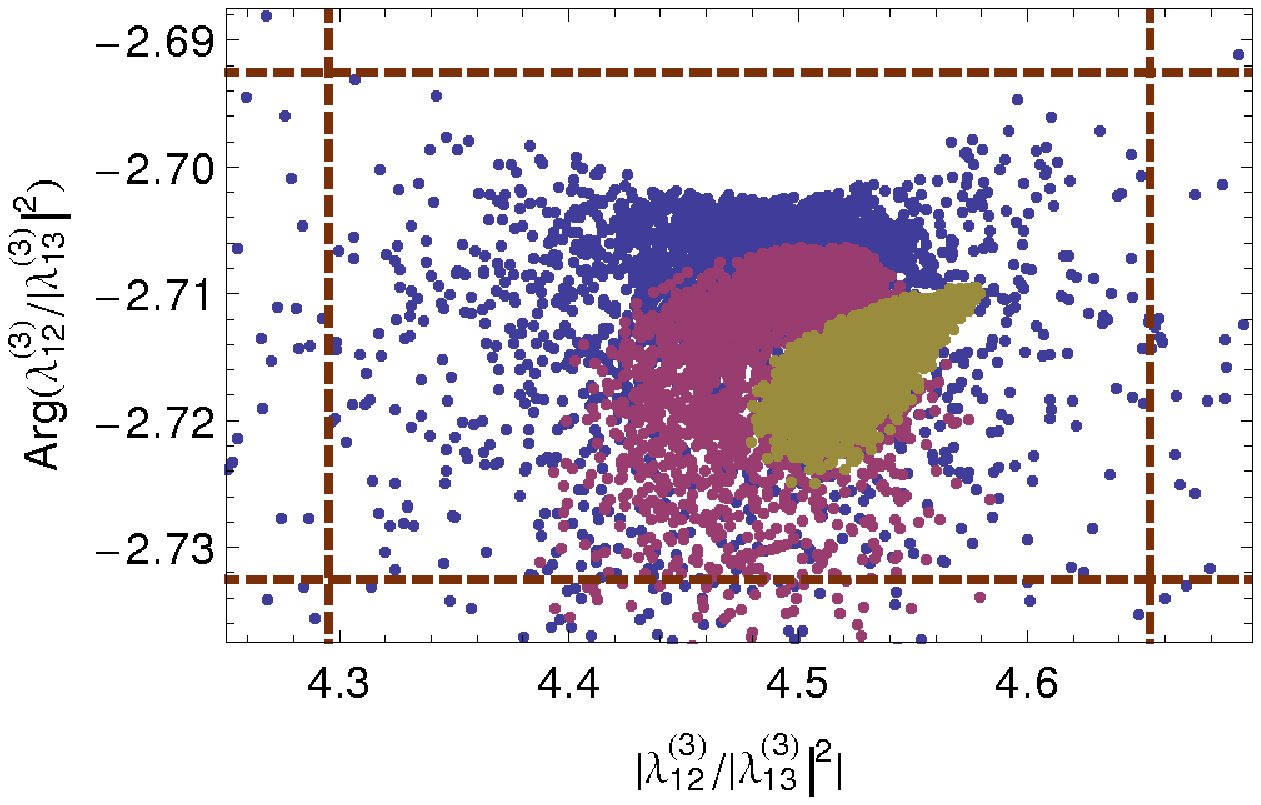}
\end{center}
\caption{The two approximate RGE-invariant ratios, evaluated at $M_{\rm SUSY}$. The dashed lines correspond to our estimated theoretical uncertainty. We show results for $x_L=2,\,1,\,0.5$ in brown, magenta and blue, respectively.} 
\label{fig:invariant}
\end{figure}

The result of evolving these ratios down to the electroweak scale is shown in Figure~\ref{fig:invariant}. Different initial values for $s_L$ are shown in different colours. The main conclusion is that in most cases the ratios are preserved within their theoretical errors. In general, the largest the initial value of $s_L$, the better the ratios are preserved.

Given the fact that the $U(2)^3$ relations are maintained at all scales, the stability when running $\lambda^{(3)}_{23}$ implies that both $\lambda^{(3)}_{12}$ and $\lambda^{(3)}_{13}$ also have a stable behaviour when solving the RGEs. This means that the virtues of the framework are preserved at all scales.

\section{Conclusions}

There exists a tension between the solution of the hierarchy problem, the current LHC bounds on sparticle masses, and the use of MFV. Furthermore, there is a slight tension between CPV observables in the $K$ and $B_d$ sectors, which MFV cannot solve. This encourages the use of other frameworks with suppression properties similar to those of MFV, but with the capacity of solving all of these tensions.

In this work, we have reviewed a $U(2)^3$ framework in SUSY, originally proposed in~\cite{Barbieri:2011ci}. We wrote the main ingredients of this framework, and explored the regions in the $m_{\tilde g}-m_{\tilde b}$ parameter space where the flavour tension would be solved. We showed that current direct search bounds constrain significantly this parameter space, forcing us to consider regions which require slightly large $\ord{1}$ parameters.

We also presented several features of this framework when defined at a high scale~\cite{Blankenburg:2012ah}. We considered the definition of the initial conditions, the evolution of the mixing, and the preservation of relations between contributions to observables in the $K$, $B_d$ and $B_s$ sectors. We defined a benchmark point and studied the parameter space around it, finding it strongly constrained. However, we also found that the mixing was rather stable during the RGE evolution, and that the $U(2)^3$ relations between different observables were preserved throughout the running.

\ack
The author would like to thank the organizers of the DISCRETE 2012 conference. The author also acknowledges support from the grants Generalitat Valenciana VALi+d, Spanish MINECO FPA 2011-23596 and the Generalitat Valenciana PROMETEO - 2008/004, and would like to thank Gianluca Blankenburg, who collaborated in the main topic of this talk.

\end{document}